\documentclass[journal=nano letters,manuscript=article]{achemso}

\usepackage{chemformula} % Formula subscripts using \ch{}
\usepackage[T1]{fontenc} % Use modern font encodings
\usepackage{upgreek}
\usepackage{graphicx}  
\usepackage[section]{placeins}
\usepackage{changes} 
\usepackage{color} 
\usepackage{hyperref}

\author{Ji-Yang Zhou}
\affiliation{CAS Key Laboratory of Quantum Information, University of Science and Technology of China, Hefei 230026, China}
\alsoaffiliation{CAS center for Excellence in Quantum Information and Quantum Physics, University of Science and Technology of China, Hefei 230026, China.}

\author{Qiang Li}
\affiliation{CAS Key Laboratory of Quantum Information, University of Science and Technology of China, Hefei 230026, China}
\alsoaffiliation{CAS center for Excellence in Quantum Information and Quantum Physics, University of Science and Technology of China, Hefei 230026, China.}

\author{Zhi-He Hao}
\affiliation{CAS Key Laboratory of Quantum Information, University of Science and Technology of China, Hefei 230026, China}
\alsoaffiliation{CAS center for Excellence in Quantum Information and Quantum Physics, University of Science and Technology of China, Hefei 230026, China.}

\author{Wu-Xi Lin}
\affiliation{CAS Key Laboratory of Quantum Information, University of Science and Technology of China, Hefei 230026, China}
\alsoaffiliation{CAS center for Excellence in Quantum Information and Quantum Physics, University of Science and Technology of China, Hefei 230026, China.}
\alsoaffiliation{Hefei National Laboratory, University of Science and Technology of China, Hefei 230088, China}

\author{Zhen-Xuan He}
\affiliation{CAS Key Laboratory of Quantum Information, University of Science and Technology of China, Hefei 230026, China}
\alsoaffiliation{CAS center for Excellence in Quantum Information and Quantum Physics, University of Science and Technology of China, Hefei 230026, China.}

\author{Rui-Jian Liang}
\affiliation{CAS Key Laboratory of Quantum Information, University of Science and Technology of China, Hefei 230026, China}

\author{Liping Guo}
\affiliation{Key Laboratory of Artificial Micro- and Nano-structures of Ministry of Education and School of Physics and Technology, Wuhan University, Wuhan 430072, China}

\author{Hao Li}
\affiliation{State Key Laboratory of Functional Materials for Informatics, Shanghai Institute of Microsystem and Information Technology, Chinese Academy of Sciences, Shanghai 20050, China}

\author{Lixing You}
\affiliation{State Key Laboratory of Functional Materials for Informatics, Shanghai Institute of Microsystem and Information Technology, Chinese Academy of Sciences, Shanghai 20050, China}

\author{Jian-Shun Tang}
\affiliation{CAS Key Laboratory of Quantum Information, University of Science and Technology of China, Hefei 230026, China}
\alsoaffiliation{CAS center for Excellence in Quantum Information and Quantum Physics, University of Science and Technology of China, Hefei 230026, China.}
\alsoaffiliation{Hefei National Laboratory, University of Science and Technology of China, Hefei 230088, China}

\author{Jin-Shi Xu}
\email{jsxu@ustc.edu.cn}
\affiliation{CAS Key Laboratory of Quantum Information, University of Science and Technology of China, Hefei 230026, China}
\alsoaffiliation{CAS center for Excellence in Quantum Information and Quantum Physics, University of Science and Technology of China, Hefei 230026, China.}
\alsoaffiliation{Hefei National Laboratory, University of Science and Technology of China, Hefei 230088, China}

\author{Chuan-Feng Li}
\email{cfli@ustc.edu.cn}
\affiliation{CAS Key Laboratory of Quantum Information, University of Science and Technology of China, Hefei 230026, China}
\alsoaffiliation{CAS center for Excellence in Quantum Information and Quantum Physics, University of Science and Technology of China, Hefei 230026, China.}
\alsoaffiliation{Hefei National Laboratory, University of Science and Technology of China, Hefei 230088, China}

\author{Guang-Can Guo}
\affiliation{CAS Key Laboratory of Quantum Information, University of Science and Technology of China, Hefei 230026, China}
\alsoaffiliation{CAS center for Excellence in Quantum Information and Quantum Physics, University of Science and Technology of China, Hefei 230026, China.}
\alsoaffiliation{Hefei National Laboratory, University of Science and Technology of China, Hefei 230088, China}

\title{Plasmonic-enhanced bright single spin defects in silicon carbide membranes}
\abbreviations{IR,NMR,UV}

\begin{document}

\begin{abstract}
Optically addressable spin defects in silicon carbide (SiC) have emerged as attractable platforms for various quantum technologies. However, the low photon count rate significantly limits their applications. We strongly enhanced the brightness by 7 times and spin-control strength by 14 times of single divacancy defects in 4H-SiC membranes using surface plasmon generated by gold film coplanar waveguides. The mechanism of the plasmonic-enhanced effect is further studied by tuning the distance between single defects and the surface of the gold film. A three-energy-level model is used to determine the corresponding transition rates consistent with the enhanced brightness of single defects. Lifetime measurements also verified the coupling between defects and surface plasmons. Our scheme is low-cost, without complicated microfabrication and delicate structures, which is applicable for other spin defects in different materials. This work would promote developing spin defect-based quantum applications in mature SiC materials.


\end{abstract}
Defects in solid-state materials with optically addressable spins have played significant roles in various quantum applications, such as distributed quantum computing~\citep{jiang2007distributed,van2016path,awschalom2018quantum}, quantum sensing~\citep{lin2021temperature,yan2018coherent,wang2022magnetic}, and quantum networking~\citep{bernien2013heralded,togan2010quantum,hensen2015loophole,xu2021network}. Recently, silicon carbide (SiC) spin defects have attracted much interest. SiC is a mature semiconductor with many desirable properties, including inch-wafer growth and well-developed fabrication technologies. Several bright single-photon emitters\citep{wang2018bright,li2019stable} and spin defects\citep{koehl2011room,widmann2015coherent,li2022room,he2022maskless,
wang2020coherent,zhou2021experimental} have been identified in SiC. Among them, the divacancy defects in SiC, losing one silicon atom and an adjacent carbon atom, have outstanding performance with infrared fluorescence\citep{yan2020room,somogyi2012near,kraus2017three} and room-temperature spin controllability\citep{klimov2015quantum,lin2022anti,yan2021room}. The divacancies have long spin coherence times\citep{christle2015isolated}, narrow optical linewidth down to lifetime limit\citep{christle2017isolated,anderson2019electrical}, and the ability to resist the photoionization effect\citep{4HSiC:VsiVc:Vsi:ensemble:OCC:01,wang2021optical}. One of the divacancy defects, named the PL6 defect, is identified as the divacancy configurations adjacent to stacking faults\citep{4HSiC:PL567:Stable:Theory:Gali}, showing high readout contrast and bright single photon count rate up to 150-kilo counts per second\citep{li2022room}, which are comparable to that of nitrogen-vacancy centers in diamond\citep{gruber1997scanning}.
Despite the great achievement of divacancy defects, their application to quantum information tasks would benefit from higher quantum efficiency and photoluminescence (PL) intensity. 
To further enhance the brightness of spin defects in SiC, many enhancement schemes can be implemented, including the solid-immersion lens\citep{widmann2015coherent,zhang2021high,bernien2012two}, nano-pillars\citep{momenzadeh2015nanoengineered,radulaski2017scalable}, bulls-eye\citep{andersen2018hybrid,li2015efficient,zheng2017chirped}, photonic crystal waveguides\citep{babin2022fabrication,bracher2015fabrication,li2015coherent,
crook2020purcell,wang2020cavity}, and fiber cavities\citep{riedel2017deterministic,ruf2021resonant,haussler2019diamond}. However, many challenging problems still need to be addressed, such as complicated microstructures and alignment between defects and structures, leading to unsatisfactory enhancement rates.

In this work, we demonstrate a conspicuous enhancement of single PL6 spin defects in 4H-SiC membranes using surface plasmon at room temperature. Surface plasmons are localized collective oscillations generated by coupled electromagnetic waves and free electrons on metal surfaces, which can speed up radiative and nonradiative decay rates \citep{gao2021high,mendelson2022coupling,yanagimoto2021purcell,li2017plasmon}. It can provide broadband enhancement covering the whole PL spectral range of the divacancy defect. We fabricated gold film coplanar waveguides on silicon substrates and pasted 4H-SiC membranes on the substrates for both PL enhancement and spin coherent control. The brightness of single PL6 defects was enhanced by a factor of 7. The saturating count rate exceeded one mega count per second (Mcps) collected by an oil objective lens with a numerical aperture (NA) of 1.3. The distance between the defect and the gold film was carefully tuned to obtain the optimal plasmonic enhancement. Besides, the optically detected magnetic resonance (ODMR) signal still maintained high readout contrast within the effect of the surface plasmon. The Rabi frequency was accelerated by a maximum of 16 times, approaching 70 megahertz (MHz). The spin coherence time and the inhomogeneous spin-dephasing time on the enhanced sample are similar to that in the bulk material. Furthermore, the enhancement effect was revealed by fitting decay rates of a three-energy-level model, which indicates a shortening of the excited state lifetime and agrees well with the enhanced saturating counts. To verify the fitting results, we also directly measured the non-resonant lifetimes of both plasmon-enhanced and non-plasmon-enhanced single PL6 defects. The results were consistent with the fitting results, confirming the coupling between defects and surface plasmons. Our scheme, without complicated micro-structures and processing technologies, is of practical significance and can be conveniently applied to other types of spin defects in various materials.
%%%%%%%%%%%%%%%%%%%%%%%%%%%%%%%%%%%%%%%%%%%%%%%%%%%%%%%%%%%%%%%%%%%%%%%%%%%%%%%%%%%%%%%%%%%%%%%%%%%%%%%%%%%%%%%%%%%%%%%%%%%%%%%%
%Fig.1
%%%%%%%%%%%%%%%%%%%%%%%%%%%%%%%%%%%%%%%%%%%%%%%%%%%%%%%%%%%%%%%%%%%%%%%%%%%%%%%%%%%%%%%%%%%%%%%%%%%%%%%%%%%%%%%%%%%%%%%%%%%%%%%%
\begin{figure}[!htb]
\begin{center}
\includegraphics[width=1\columnwidth]{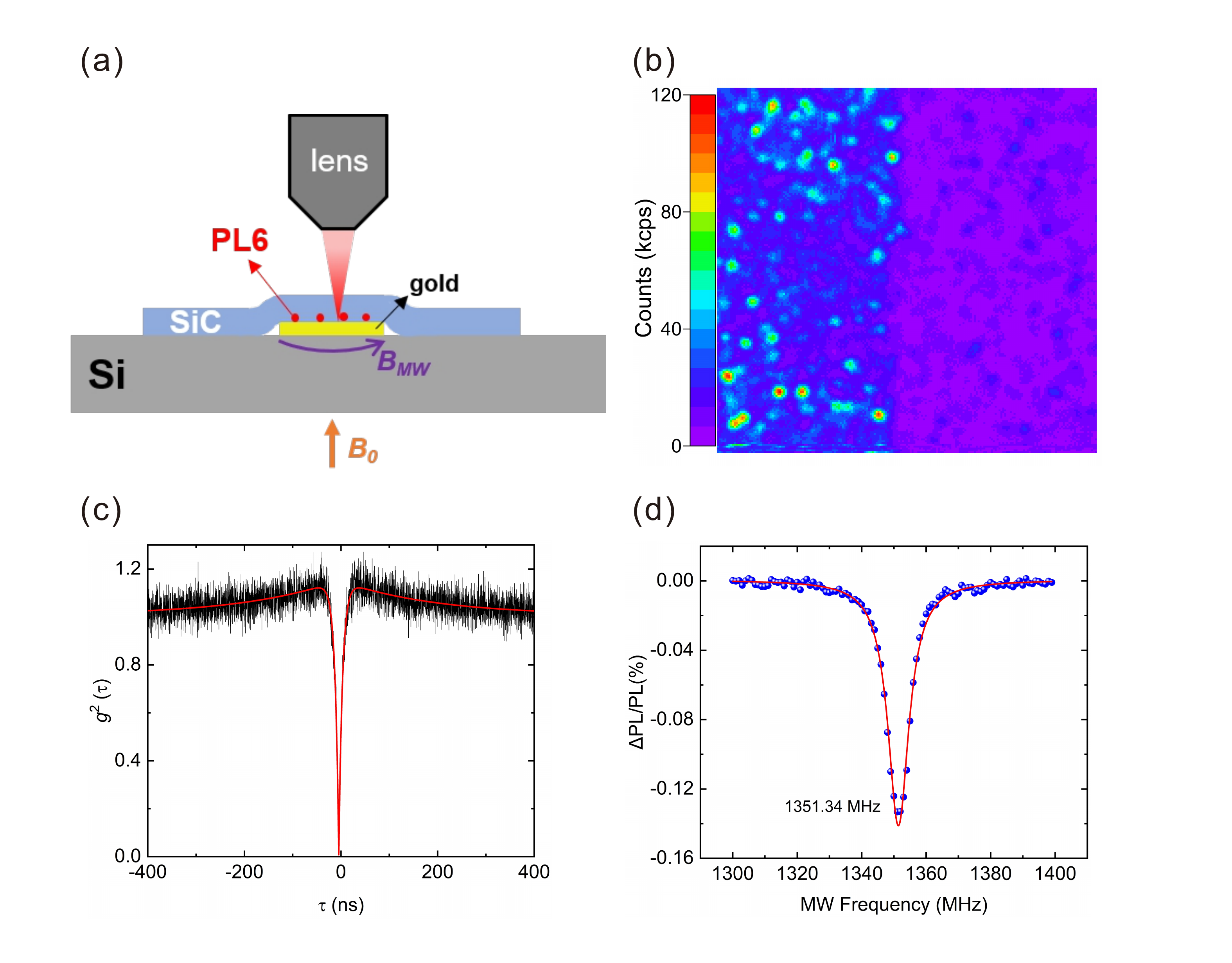}
\caption{(a) The sketch of the device’s structure. The gold film  coplanar waveguide with a thickness of $210$ nm is coated on a silicon substrate. The ion-implanted 4H-SiC membrane is overturned and pasted on the substrate by Van der Waals force so that the shallow spin defects can be close to the gold coplanar surface. The microwave (MW) transmits through the waveguide with the magnetic field ($B_{MW}$) to manipulate spins. The applied external static magnetic field ($B_{0}$) is along the c-axis. An objective lens  is to focus the pump laser beam and collect the PL. (b) A confocal scanning image of spin defects above (left, bright) and beside (right, dark) the gold coplanar waveguide. The PL enhancement is obtained with the gold film. (c) The second-order intensity correlation $g^2(\tau)$ measurement of a single PL6 defect. The red line is the fitting given $g^{2}(0)=0.008\pm$0.002, which is significantly smaller than 0.5 verifying the single-photon emitter. (d) The ODMR spectrum of the single PL6 defect without the external static magnetic field.}
\label{scheme}
\end{center}
\end{figure}

The enhancement device used in this work comprises one 4H-SiC membrane and a piece of silicon substrate coated with gold coplanar waveguides, shown in Fig.~\ref{scheme}a. The SiC membrane was fabricated by mechanically grinding and chemically mechanically polishing a SiC wafer. The SiC epitaxial layer (30 $\upmu$m thick) on the wafer's substrate (350 $\upmu$m thick) was finally thinned to 12.5 $\upmu$m. The membrane was then implanted by the N${_2^+}$ ions with 30-kilo electron volts energy and a dose of $1 \times 10^{12}$ cm$^{-2}$. The sample was annealed under 1050 ${^\circ}$C (2 hours) to generate single PL6 defects efficiently. The Stopping and Range of Ions in Matter (SRIM) simulation results show that the divacancies are distributed centrally 15 nm below the surface (see supplementary information (SI) for more details). Using photolithography and evaporation coating, we fabricated 210-nm thick gold coplanar waveguides on silicon substrates. The SiC membrane was overturned and pasted on the silicon substrate by Van der Waals force, with the implanting side facing the gold waveguide. So the near-surface PL6 defects can be close enough to the gold surface and come into the effective range of the surface plasmon. We used a homebuilt confocal scanning system to pump the sample with a continuous-wave 914 nm laser and collected the PL with the same objective. A superconducting single-photon detector recorded the PL intensity. A Hanbury-Brown and Twiss (HBT) interferometer was used to measure the second-order intensity correlations. The transmitted microwave through the gold waveguide manipulated the defect spins. An external magnetic field ($B_{0}$) along the c-axis can be added to the sample.

The confocal scanning image shown in Fig.~\ref{scheme}b drastically displays the different brightness between the defects above the gold film (the left side, bright area) and beside the gold film (the right side, shaded area). Fig.~\ref{scheme}c shows the HBT measurement of one single PL6 defect. The measured second-order intensity correlation $g^2(\tau)$ was fitted by the red line giving $g^{2}(0)=$0.008$\pm$0.002 at the delay time $\tau=0$, which is significantly smaller than 0.5 and confirms an excellent single-photon emitter. Fig.~\ref{scheme}d shows the ODMR spectrum of a single PL6 defect without static external magnetic fields. The resonant ODMR signal centering at 1351.34 MHz agrees with the previous results\citep{li2022room}.

%%%%%%%%%%%%%%%%%%%%%%%%%%%%%%%%%%%%%%%%%%%%%%%%%%%%%%%%%%%%%%%%%%%%%%%%%%%%%%%%%%%%%%%%%%%%%%%%%%%%%%%%%%%%%%%%%%%%%%%%%%%%%%%%
%Fig.2
%%%%%%%%%%%%%%%%%%%%%%%%%%%%%%%%%%%%%%%%%%%%%%%%%%%%%%%%%%%%%%%%%%%%%%%%%%%%%%%%%%%%%%%%%%%%%%%%%%%%%%%%%%%%%%%%%%%%%%%%%%%%%%%%
\begin{figure}[!htb]
\begin{center}
\includegraphics[width=1\columnwidth]{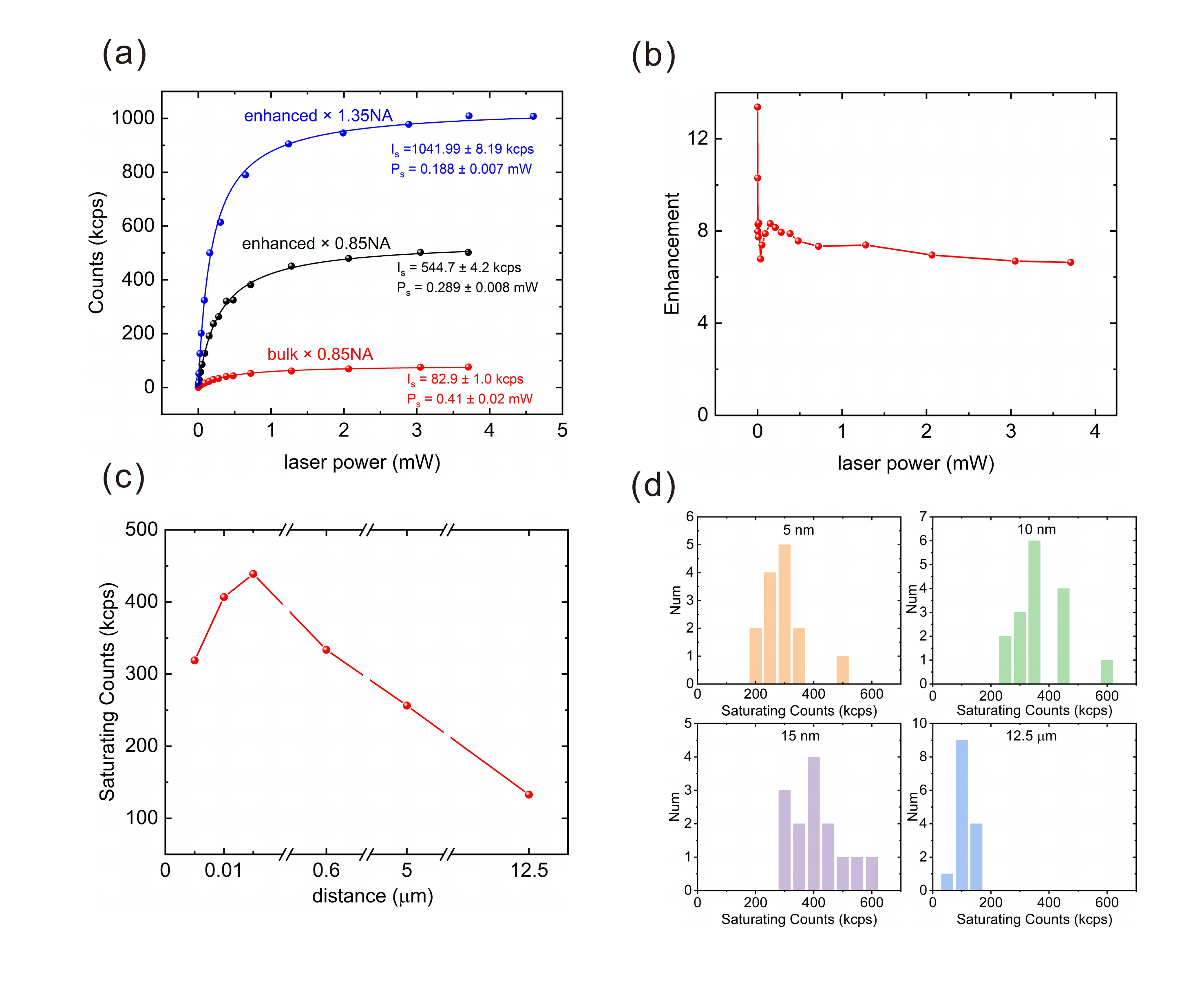}
\caption{Characterization of the enhanced PL of single PL6 defects. (a) Saturating counts ($I_{s}$) comparison between single PL6 defects in a 4H-SiC membrane on the gold coplanar waveguide and a bulk 4H-SiC. The highest count collected by an oil objective with NA = 1.3 exceeds 1 Mcps. The saturating laser power ($P_{s}$) decreases with the increased saturating counts. (b) The enhancement ratio of PL counts collected by the objective with 0.85 NA at different laser power. The ratio is calculated by dividing the intensity from the bulk sample by the membrane sample's. The ratio reaches 14 when the laser power is low. (c) The saturating counts of single PL6 defects at different distances between the defects and the surface of the gold coplanar waveguide. The results are the statistic of saturating counts from dozens of single PL6 defects. (d) The saturating counts of dozens of PL6 defects are classified by taking 50-kilo counts per second (kcps) as the interval for distances of 5 nm, 10 nm, 15 nm, and 12.5 $\upmu$m, respectively. The counts at 5 nm shift left comparing with the results of 15 nm, and the distribution width at 12.5 $\upmu$m is narrower than others.}
\label{enhanced_f}
\end{center}
\end{figure}

We compared the saturating counts of single PL6 defects in different situations to show the enhancement of the gold coplanar waveguide. The red and black dots in Fig.~\ref{enhanced_f}a represent the photon counts of a single PL6 defect in a 4H-SiC bulk (the 30 $\upmu$m epitaxial layer on a 350 $\upmu$m substrate) and the 4H-SiC membrane in Fig.~\ref{scheme}a, respectively. Both PL counts were collected by a 0.85 NA objective. The red and black lines represent the corresponding fittings using the function of $I$ = $I{_s}P/(P+P{_s})$, where $I$ is the photon counts, $I{_s}$ is the saturating count, $P$ is the laser power, and $P{_s}$ is the saturating laser power. The saturating count of the single defect in the membrane is $I_{s}=544.7\pm4.2$ kcps with the saturating laser power $P_{s}=0.289\pm0.008$ mW, compared to $I_{s}=82.9\pm1.0$ kcps and $P_{s}=0.41\pm0.02$ mW in the bulk material. The saturating count increases 6.5 times, and the corresponding saturating laser power decreases with the assistance of plasmon enhancement.

When the NA increased from 0.85 to 1.3 (oil objective), the enhanced saturating count increased to $I_{s}=1041.99\pm8.19$ kcps, and the saturating power was further reduced to $P_{s}=0.188\pm0.007$ mW. The enhanced photon counts are stable even at high pump laser power (see SI for the measurement of photon stability). Figure \ref{enhanced_f}b shows the fluorescence enhancement ratio at corresponding laser power between the enhanced and the bulk counts collected by the 0.85 NA objective. The enhanced ratio can reach 14 when the laser power is low. 

To demonstrate the distance effect on the PL enhancement, we systematically studied the saturating counts of single PL6 defects that vary with the distance between defects and the gold film. According to the SRIM simulation, the implanted single defects are centered at about 15 nm below the surface. The membrane was closely bonded on the gold film using the Van der Waals force, which was confirmed by the zero-order interference fringes (see SI for more details). As a result, the distance is equal to that between the defect and the gold film without an air gap. To further decrease the distance, we etched the implanted surface of membranes using reactive ion etching (RIE) by 5 nm and 10 nm, respectively\citep{li2019nanoscale} (see SI for etching details). So the corresponding distances between defects and the gold film at the tightly bonded region became 10 nm and 5 nm, respectively. On the other hand, the larger distance was obtained using the loosely pasted membrane, verified by high-order interference fringes or even no fringes. The thickness was then measured with a step profiler. By subtracting the thickness at the tightly bonded region, we extracted the distance between the defects and the gold film in the loosely pasted areas (see SI for more details). The saturating counts of single PL6 defects at distances 0.6 $\upmu$m and 5 $\upmu$m were chosen to measure. For the distance of 12.5 $\upmu$m, we pasted the as-prepared membrane without overturning and measured. We summarized the experimental statistics saturating counts at each distance in Fig.~\ref{enhanced_f}c. The maximal saturating count was obtained at a distance of 15 nm. The saturating counts decrease whether the distance is larger or smaller than 15 nm, demonstrating a strong distance dependence of the plasmon enhancement, which is consistent with previous work\citep{gao2021high}.

Considering the implanted single PL6 defects distribution, the saturating counts in Fig.~\ref{enhanced_f}c were averaged out from dozens of PL6 defects. The statistics of saturating counts at each distance are shown in Fig.~\ref{enhanced_f}d. We took 50-kilo counts as the interval to classify dozens of saturating counts. When single PL6 defects are close to the gold film within the effect of surface plasmon (5 nm, 10 nm, and 15 nm distances), the distribution of saturating counts is obviously broader than that without the plasmon enhancement effect (12.5 $\upmu$m distance). It is because the surface plasmon enhancement is sensitive to the defect distance within nanometer scales. The results at distances 0.6 $\upmu$m and 5 $\upmu$m are shown in SI, which also have narrower distribution widths.
%%%%%%%%%%%%%%%%%%%%%%%%%%%%%%%%%%%%%%%%%%%%%%%%%%%%%%%%%%%%%%%%%%%%%%%%%%%%%%%%%%%%%%%%%%%%%%%%%%%%%%%%%%%%%%%%%%%%%%%%%%%%%%%%
%Fig.3
%%%%%%%%%%%%%%%%%%%%%%%%%%%%%%%%%%%%%%%%%%%%%%%%%%%%%%%%%%%%%%%%%%%%%%%%%%%%%%%%%%%%%%%%%%%%%%%%%%%%%%%%%%%%%%%%%%%%%%%%%%%%%%%%
\begin{figure}[!htb]
\begin{center}
\includegraphics[width=0.7\columnwidth]{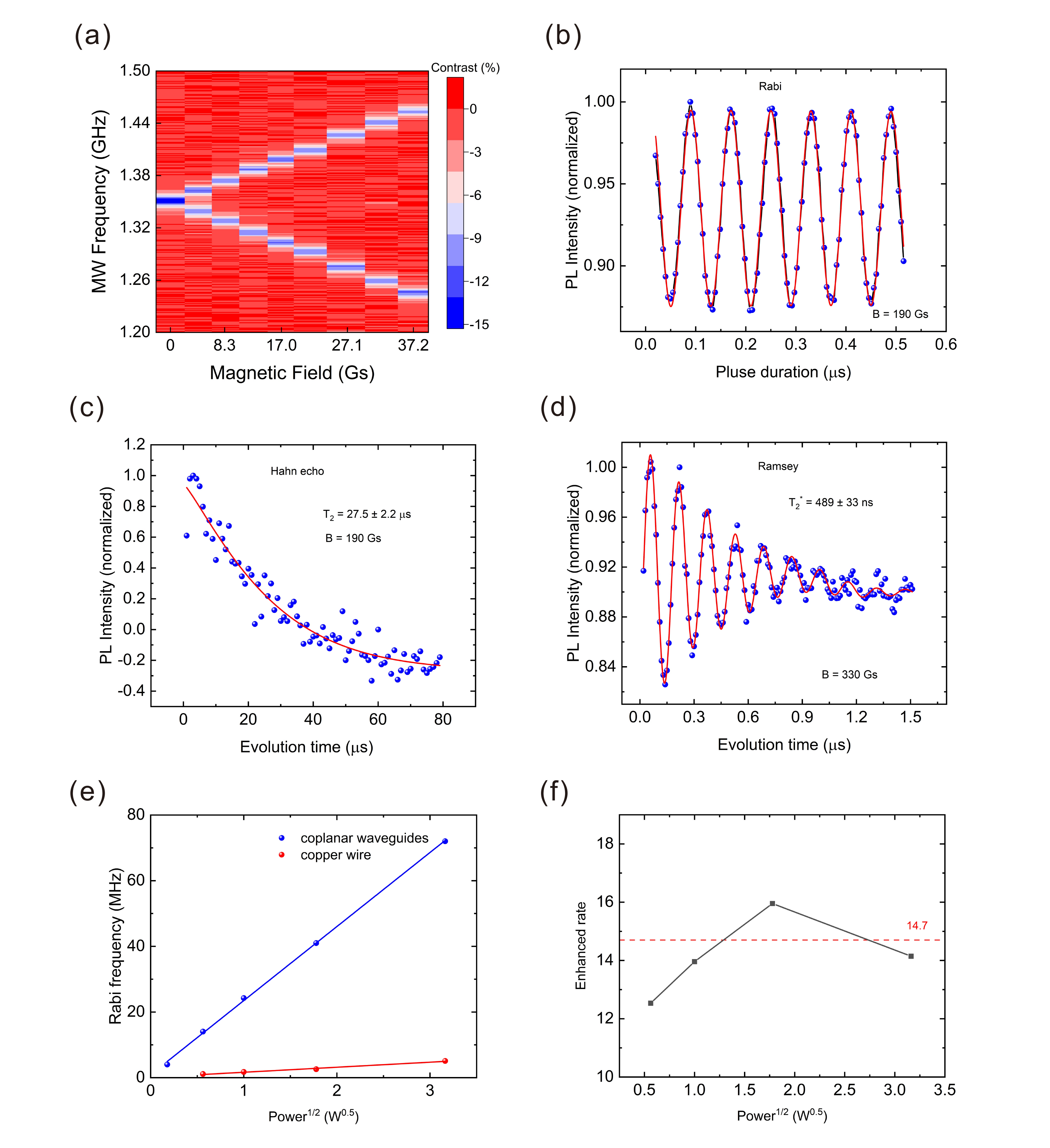}
\caption{Spin properties of single PL6 defects in 4H-SiC membranes on the gold coplanar waveguides. (a) The CW-ODMR spectrum under different static magnetic fields. The linearly split slopes are ${\pm}$ 2.8 MHz/Gs. (b) Rabi oscillation at a static magnetic field of 190 Gs. (c) Hahn echo measurement at a magnetic field of 190 Gs. The homogeneous spin coherence time ${T_2}$ is 27.5 ${\pm}$ 2.2 $\upmu$s from the fitting. (d) Ramsey measurement at a magnetic field of 190 Gs. The inhomogeneous spin-dephasing time ${T_2^*}$ is 489 ${\pm}$ 33 ns. (e) Rabi frequencies of single PL6 defects at different microwave powers transmitted by the gold coplanar waveguide (blue dots) and a copper wire (red dots), respectively. The Rabi frequency is strongly enhanced with the gold film compared that with the copper wire. Both Rabi frequencies obey the linear relationship with the square root of the microwave power, which are fitted by the bule and red lines, respectively. (f) The corresponding enhanced ratio of Rabi frequencies at different microwave powers. The gold coplanar waveguide gives rise to a maximal 16 times enhancement of Rabi frequencies. The red dotted line of 14.7 is the ratio between the theoretical fittings in (e).}
\label{enhanced_mw}
\end{center}
\end{figure}
\FloatBarrier 

Coherent spin controls are essential for quantum information processing. By implementing microwave pulses, we characterized the spin properties of single PL6 defects in the membrane with a distance of 15 nm from the gold coplanar waveguides, which is within the effect of the surface plasmon. Fig.~\ref{enhanced_mw}(a) shows the continuous-wave ODMR spectrum of a single PL6 defect as a function of the external magnetic field. The slopes of the splitting are ${\pm}$ 2.8 MHz/Gs due to the Zeeman effect, which agrees well with previous results\citep{li2022room,wang2020optimization}. Figs.~\ref{enhanced_mw}(b), (c), and (d) demonstrate the corresponding Rabi, spin echo, and Ramsey oscillations with an external magnetic field of 190 Gs. The corresponding Rabi oscillation frequency is 12.5 MHz. The homogeneous spin coherence time ${T_2}$ was deduced to be 27.5 ${\pm}$ 2.2 $\upmu$s, and the inhomogeneous spin dephasing time ${T_2^*}$ was deduced to be 489 ${\pm}$ 33 ns. These typical coherence times are consistent with previous results measured in bulk materials\citep{li2022room}, illustrating no degradation caused by the surface plasmon effect. Moreover, compared with the case using a 50 $\upmu$m diameter copper wire to supply the microwave, the gold coplanar waveguide has higher radiation efficiency, and the magnetic field component can be more perpendicular to the c-axis of PL6 defects. As a result, the Rabi frequency can get an enormous acceleration. Fig.~\ref{enhanced_mw}(e) compares the Rabi frequencies in these two cases as a function of microwave power. The Rabi oscillation frequency using the gold coplanar waveguides is larger than that using the copper wire, reaching as high as 70 MHz. Both Rabi frequencies have linear relationships with the square root of microwave power. The fitting slops are 22.56 for the gold waveguides (blue line) and 1.53 for the copper wire (red line), respectively. The enhanced Rabi frequency ratio between both cases at corresponding microwave power is shown in Fig.~\ref{enhanced_mw}(f). The gold coplanar waveguide provides up to 16-fold enhancement of the Rabi oscillation frequency. The red dashed line of 14.7 is the ratio of the linear fitting slopes.

%%%%%%%%%%%%%%%%%%%%%%%%%%%%%%%%%%%%%%%%%%%%%%%%%%%%%%%%%%%%%%%%%%%%%%%%%%%%%%%%%%%%%%%%%%%%%%%%%%%%%%%%%%%%%%%%%%%%%%%%%%%%%%%%
%Fig.4
%%%%%%%%%%%%%%%%%%%%%%%%%%%%%%%%%%%%%%%%%%%%%%%%%%%%%%%%%%%%%%%%%%%%%%%%%%%%%%%%%%%%%%%%%%%%%%%%%%%%%%%%%%%%%%%%%%%%%%%%%%%%%%%%
\begin{figure}[!htb]
\begin{center}
\centerline{\includegraphics[width=01.2\columnwidth]{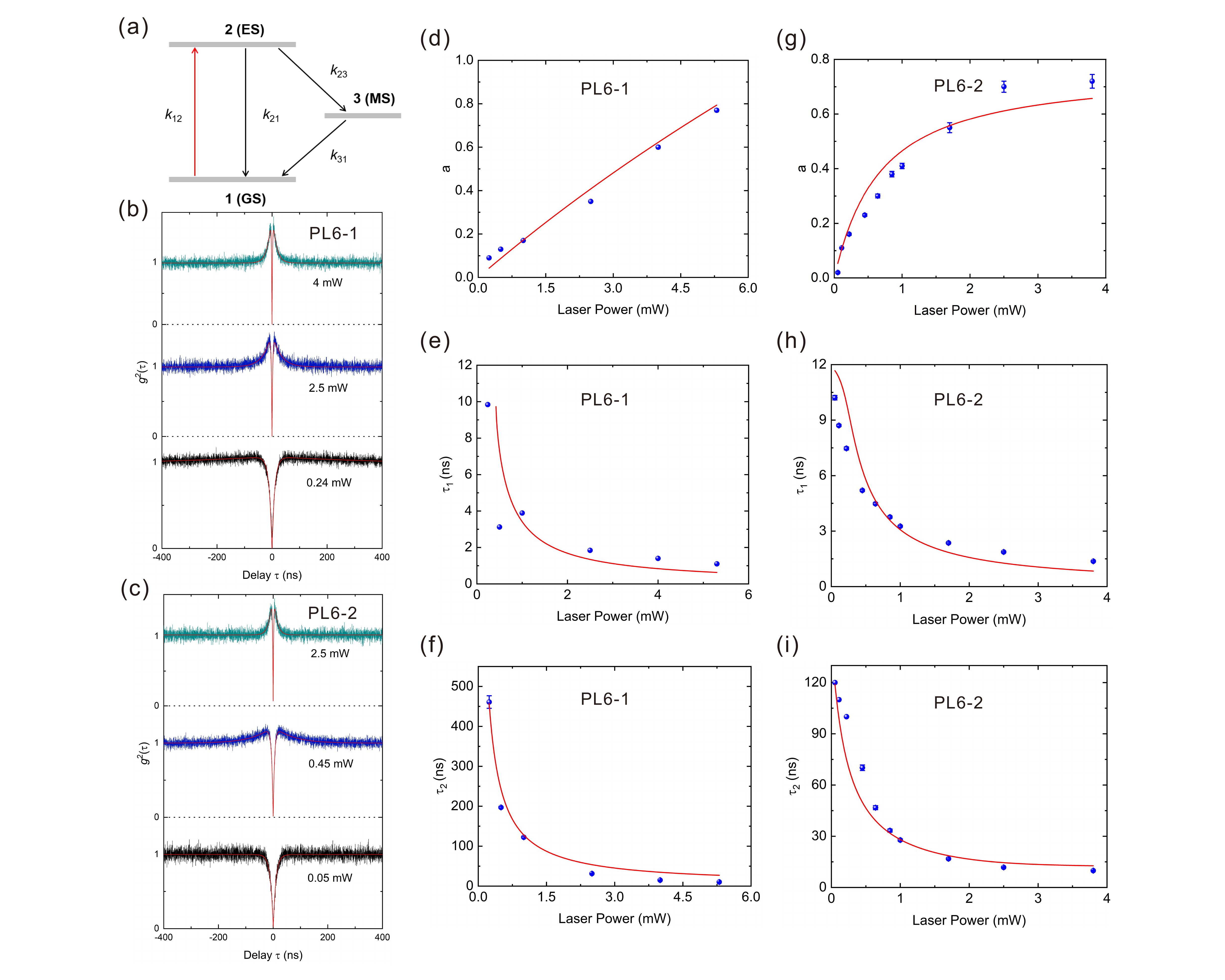}}
\caption{Energy level analysis of two single PL6 defects. The defects PL6-1 and PL6-2 are found in samples with distances of 12.5 $\upmu$m and 5 nm from the gold waveguides' surface, respectively. (a) The three-level model of PL6 defects is composed of the ground state (GS), excited state (ES), and metastable state (MS). $k_{ij}$ represent the transition rates between the levels $i$ and $j$. (b) and (c) Laser-power dependent second-order intensity correlations of PL6-1 and PL6-2, respectively. (d-f) and (g-i) The fitting parameters $a$, $\tau_1$, and $\tau_2$ in the correlation function $g^2(\tau)$ as a function of laser power for PL6-1 and PL6-2, respectively. Error bars represent the standard deviations of the corresponding $g^2(\tau)$ fittings.}
\label{three_level}
\end{center}
\end{figure}
\FloatBarrier

To deeply understand the mechanism of the surface plasmon enhancement, we measured laser power-dependent intensity correlations for both enhanced and non-plasmon-enhanced single PL6 defects. We used a three-level model to characterize the energy levels of PL6 defects, shown in Fig.~\ref{three_level}(a). The three energy levels, 1, 2, and 3, correspond to the ground state (GS), excited state (ES), and metastable state (MS). $K_{ij}$ represents the transition rate from level $i$ to level $j$. Fig.~\ref{three_level}(b) and (c) show three representative laser-power dependent intensity correlations of a single PL6 defect without plasmon enhancement (PL6-1) and an enhanced single PL6 defect (PL6-2), respectively. The saturating counts of these two defects are 161.85 kcps and 405.48 kcps, respectively, as shown in SI.  

With the increase of laser power, the bunching effect becomes clear for both defects, implying the existence of a metastable state of the PL6 defects. We corrected the raw data ${g^2_{raw}(\tau)}$ to remove the influence of the background by the function $g^2(\tau)=[g^2_{raw}(\tau)-(1-\rho^2)]/\rho^2$,
where $\rho=s/(s+b)$. $s$ and $b$ represent the signal and background counts, respectively. The background-corrected intensity functions are fitted by $g^2(\tau)=1-(1+a)e^{|\tau|/\tau_1}+ae^{|\tau|/\tau_2}$, where $a$, $\tau_1$ and $\tau_2$ are power-dependent fitting parameters. Fig.~\ref{three_level}(d-f) and (g-i) show the fitting parameters $a$, $\tau_1$, and $\tau_2$ of PL6-1 and PL6-2 as a function of the laser power, respectively. The parameter $a$ reveals the bunching effect, $\tau_1$ represents the anibunching decay time, and $\tau_2$ reflects the relatively long bunching decay time\citep{fuchs2015engineering}. These three parameters were fitted using the following equations based on the three-level model\citep{wang2018bright,fuchs2015engineering}:
\begin{equation}\label{a}
a=\frac{1-\tau_2k_{31}}{k_{31}(\tau_2-\tau_1)},
\end{equation}
\begin{equation}\label{t1t2}
\tau_{1,2}=\frac{2}{A\pm\sqrt{A^2-4B}},
\end{equation}
where $A=k_{12}+k_{21}+k_{23}+k_{31}$, $B=k_{12}(k_{23}+k_{31})+k_{31}(k_{21}+k_{23})$. We further considered another enhanced single PL6 defect (PL6-3); the corresponding data are shown in SI. The fitting parameters for these three single PL6 defects are summarized in Table \ref{fitting_params}.

\begin{table}
  \caption{Decay rates $k_{ij}$ and saturating counts of different PL6 defects.}
  \label{fitting_params}
  \begin{tabular}{ccccc}
    \hline
    single PL6 defects  & {1/$k_{21}$} & {1/$k_{23}$} & {1/($k_{21}+k_{23}$)} & saturating counts  \\
    \hline
    PL6-1 & 200.00 ns & 15.40 ns  & 14.31 ns & 161.85 kcps  \\
    PL6-2 & 83.30 ns  & 13.87 ns  & 11.89 ns & 405.48 kcps \\
    PL6-3 & 94.34 ns  & 15.50 ns  & 13.32 ns & 341.29 kcps \\
    \hline
  \end{tabular}
\end{table}

According to the three-level model, the brightness of single PL6 defects is directly proportional to $k_{21}$. The obtained $k_{21}$ are consistent with the saturating counts for all three defects. The ratio of $k_{21}$  between PL6-2 (PL6-3) and PL6-1 is 2.4 (2.12), which agrees well with the corresponding saturating counts' ratio of 2.51 (2.11). Therefore, the enhanced PL of single PL6 defects may mainly be due to the increased transition rates between ES and GS enhanced by the surface plasmon.

\begin{figure}[H]
\centering
\includegraphics[width=1\columnwidth]{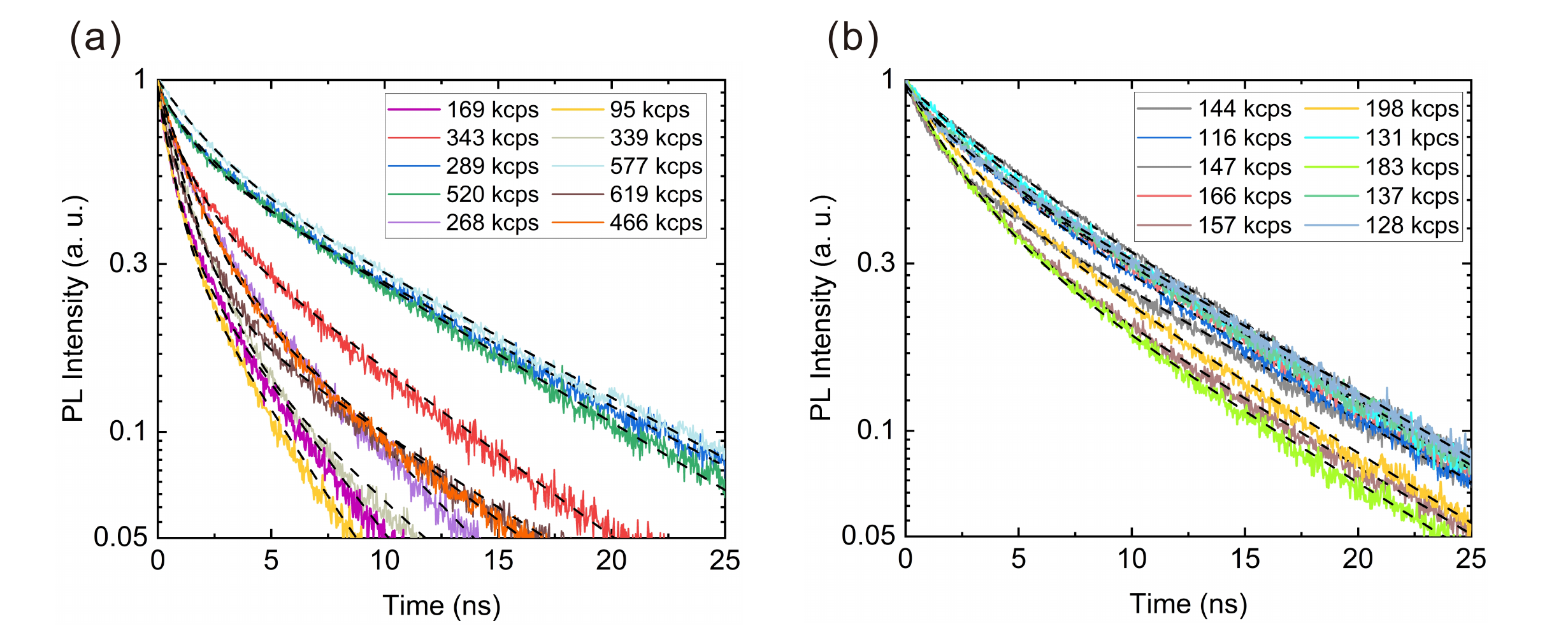}
\caption{Lifetimes of different PL6 defects with various saturating counts. The shortening of enhanced PL6's lifetime clearly verifies the coupling between defects and surface plasmon. The decay is well fitted by a biexponential decay demonstrated by the black dashed lines. (a) Lifetimes of ten enhanced PL6 centers with different saturating counts. (b) Lifetimes of ten non-plasmon-enhanced PL6 centers with a small difference in saturating counts.}
\label{lifetime}
\end{figure}

To confirm the results of the three-energy-level model that predicts the enhanced transition rates and the shortened lifetime of defects, we measured the non-resonant excited lifetime of ten enhanced single PL6 defects in the 5 nm sample and ten non-plasmon-enhanced in the 12.5 $\upmu$m sample, respectively,  shown in Fig.~\ref{lifetime}. The data were well fitted by a biexponential decay function and demonstrated an obvious shortening of enhanced defects compared with non-plasmon-enhanced defects. The lifetimes are summarized in Table \ref{lifetime_1} and \ref{lifetime_2}. We observe that the saturating counts do not monotonically change with the shortened lifetimes. Moreover, the lifetime of PL6 with the highest saturating count is not significantly reduced. As shown in Table \ref{lifetime_1}, although two PL6s have almost the same saturating counts (289 kcps and 268 kcps), the lifetimes are significantly different. Considering the distribution of PL6 defects shown in Figure S1(a) in SI, the fact that radiative outcoupling will be quenched if the emitter is close to the metal surface\citep{reineck2013distance,wang2014core,yan2017high}, and the different lifetimes mean the different distances from the gold surface of two defects. The similar counts imply the quenching effect has decreased the closer PL6's saturating counts, although it has a shorter lifetime. This quenching effect is further studied in the simulation results in SI. On the other hand, the lifetime of the PL6 with the highest counts is not significantly dropped This is because the measured non-resonant lifetime also includes the non-radiative decay rate $k_{23}$\citep{fuchs2015engineering,wang2018bright}. The radiative decay rate $k_{21}$ in Table 1 is significantly increased, which verifies that the defect is coupled with the surface plasmon as well.

\begin{table}
  \caption{Lifetimes of ten enhanced single PL6 defects with different saturating counts.}
  \label{lifetime_1}
  \begin{tabular}{c|cccccccccc}
    \hline
    saturating counts (kcps) & 169 & 343 & 289 & 520 & 268 & 95 & 339 & 577 & 619 & 466\\
    \hline
   	lifetime (ns) & 5.0 & 8.9 & 12.3 & 11.0 & 6.4 & 4.3 & 8.4 & 10.1 & 10.1 & 8.5 \\
    \hline
  \end{tabular}
\end{table}

\begin{table}
  \caption{Lifetimes of ten non-plasmon-enhanced single PL6 defects.}
  \label{lifetime_2}
  \begin{tabular}{c|cccccccccc}
    \hline
    saturating counts (kcps) & 144 & 116 & 147 & 166 & 157 & 198 & 131 & 183 & 137 & 128 \\
    \hline
   	lifetime (ns) & 11.7 & 11.1 & 11.4 & 11.3 & 11.2 & 10.8 & 11.6 & 10.7 & 11.6 & 11.4 \\
    \hline
  \end{tabular}
\end{table}
In conclusion, we constructed a simple but high-performance single-defect enhancement device by pasting 4H-SiC membranes onto a silicon substrate coated with a thin gold coplanar waveguide. The fluorescence got a 7-fold enhancement and could exceed 1 Mcps at room temperature enhanced by the surface plasmon. Since the gold waveguide has high radiative efficiency of the microwave, and the magnetic component can be perpendicular to the c-axis of single PL6 defects compared with the copper wire, the obtained spin Rabi frequency is enhanced to 16 times and reaches as high as 70 MHz. We further optimized the fluorescence enhancement effect by tuning the distance between the gold coplanar waveguide and the defects and obtained the maximal saturation counts at a distance of about 15 nm. We measured laser-power-dependent second-order intensity correlations to fit the transition rates in a three-energy-level model. The fitted parameters agree well with the experimental results, which indicates that the surface plasmon accelerates the inner transition rates of single defects and enhances the fluorescence. We further directly measured the lifetimes of both plasmon-enhanced and non-plasmon-enhanced single PL6 defects, consistent with the fitted transition rates, confirming the coupling between defects and surface plasmons. Our device, without complex fabrication technologies and structures, can conveniently migrate to other types of defects in different materials. With accelerated coherent spin control, the enhanced brightness of single PL6 defects can further improve the signal-to-noise ratio of the photoluminescence excitation and the state readout fidelity, which is helpful for quantum networking based on SiC~\citep{xu2021network}. The enhanced structure can also combine with other enhancement structures, such as photonic cavities and the bull eye structure, to further enhanced the fluorescence collection. This enhancement scheme would play significant roles in quantum applications using spin defects in solid-state materials.

\begin{acknowledgement}

This work was supported by the Innovation Program for Quantum Science and Technology (Grants No. 2021ZD0301400), the National Natural Science Foundation of China (Grants No. U19A2075, 61725504 and 11821404), Anhui Initiative in Quantum Information Technologies (Grants No. AHY060300). This work was partially carried out at the USTC center for Micro and Nanoscale Research and Fabrication. 

\end{acknowledgement}

\begin{suppinfo}
Additional details as follows: (1) The simulation results of the ion implantation and the image of the 4H-SiC membrane pasted on the silicon substrate; (2) Saturating counts of three different single PL6 defects; (3) Laser-power-dependent intensity correlations and fitting parameters of PL6-3; (4) Statical distribution of defects 0.6 $\upmu$m and 5 $\upmu$m from the gold surface, and stable count records of a single PL6 defect. (5) FDTD simulation results of transition rates for three different dipole orientations. (6) Height measurement of the etched step using an AFM. (7) Height measurement of the bonded membrane on the substrate using a step profiler.
\end{suppinfo}

\noindent
{\bf Notes}\\
\noindent
The authors declare no competing financial interest.\\

\end{document}